\documentstyle[axodraw,citesort,epsfig]{elsart}

\renewcommand{\thefootnote}{\fnsymbol{footnote}}

\newcommand{\tanbeta}{$tan\beta$ } 
\newcommand{\kltomupm}{$K_L \to \mu^+ \mu^-$}
\newcommand{\beq}{\begin{equation}}
\newcommand{\eeq}{\end{equation}}

\def\prd#1#2#3{ {\em Phys.Rev. } {\bf D#1},{ #2} (#3)}

\def\plb#1#2#3{{\em Phys. Lett. }{\bf B #1},{#2}(#3)}
\def\nuphsb#1#2#3{{\em Nucl.Phys. }{\bf B #1},{#2}(#3)}

\begin{document}

\journal{Physics Letters B}
                                                                 

\begin{frontmatter}

\title{\large \bf \boldmath Longitudinal Polarization in $K_L \to
\mu^+ \mu^-$ in MSSM with large $tan\beta$}

\author{S. Rai Choudhury \thanksref{src}}
\author{Naveen Gaur \thanksref{naveen}}
\author{Abhinav Gupta \thanksref{abh}}
\address{ Department of Physics \& Astrophysics \\
                  University of Delhi \\
                  Delhi - 110 007,India }          

\thanks[src]{E-mail : src@ducos.ernet.in}
\thanks[naveen]{E-mail : naveen@physics.du.ac.in,
pgaur@ndf.vsnl.net.in}  
\thanks[abh]{E-mail : abh@ducos.ernet.in}

\begin{abstract}
  A complete experiment on decay $K_L \to l^+ l^-$ will not only consist of
measurement of the decay rates but also lepton polarization etc. These 
additional observations will yield tests of CP invariance in these
decays. In $K_L$ and $K_S$ decays , the e mode is slower than the
$\mu$ mode by roughly $(m_e/m_\mu)^2$ \cite{sehgal1}. As well
discussed in literature \cite{herczeg} the Standard Model contribution
to the lepton polarization is of order $2 \times \sim 10^{-3}$. We
show that in MSSM with large \tanbeta and light higgs masses ($\sim 2
M_W$), the  longitudinal lepton polarization in $K_L \to \mu^+ \mu^-$
can be enhanced to a higher value, of about $10^{-2}$.  
\end{abstract}

\end{frontmatter}


\pagestyle{plain}

      The Flavor Changing Neutral Current (FCNC) decays of the K-meson
are forbidden in the lowest order in the standard Electroweak theory
but can occur through loop diagrams in higher order. In effect these 
processes thus are a deeper probe into the underlying field
theory. The amplitude for such processes can be divided into a
Short-Distance (SD) part where the quarks involved interact over a
range $\sim M_W^{-2}$ and a Long Distance (LD) part which may be
thought of as proceeding via low lying intermediate states and
particularly through resonances. Theoretically reasonable techniques
have been developed for estimating the SD - part through Operator
Product Expansion (OPE) \cite{buchalla1} and this is quite successful
in analyzing decays like $b \to s \gamma$ or $b \to s l^+ l^-$ in
regions of phase space where no resonances are involved
\cite{stone1,buras1}. For K-meson decays, resonances are nearby and
the LD part is important. Unfortunately the LD-part is quite model
dependent and thus theoretical results are relatively more uncertain
compared to the ones for B-decay.

\begin{center}
\vskip -1cm
%

\begin{picture}(200,200)(0,0)

\ArrowLine(0,100)(75,100)
\Photon(75,100)(150,130){3}{5}
\Photon(75,100)(150,70){3}{5}
\Line(150,130)(150,70)
\ArrowLine(150,130)(200,130)
\ArrowLine(150,70)(200,70)
\GCirc(75,100){5}{0}

\Text(30,120)[c]{\large $K_L$}
\Text(100,130)[c]{\large $\gamma$}
\Text(100,70)[c]{\large $\gamma$}
\Text(180,150)[c]{\large $\mu^+$}
\Text(180,50)[c]{\large $\mu^-$} 

\end{picture}


\vskip -1cm
\noindent Figure~1 :{\it The $2 \gamma$ intermediate state
contribution to the LD part of \kltomupm} 
\end{center}
\smallskip

\par The decay of $K_L$ into $\mu^+ \mu^-$, a FCNC process, is a
somewhat special one amongst all rare decays. Amongst the intermediate
states which contribute to the LD-part, the two photon state (Fig -1)
stands as the most important one \cite{sehgal1}. The absorptive part
of the amplitude, which can be computed from the known decay rate $K_L
\to 2 \gamma$ and the QED amplitude $\gamma \gamma \to \mu^+ \mu^-$,
by itself leads to a decay rate almost equal to experimental decay
rate of $K_L \to \mu^+ \mu^-$ \cite{herczeg}. The short-distance
contributions are real \cite{buchalla2} and since the rate of \kltomupm
depend on the sum squares of the absorptive and the real parts, the SD
parts become somewhat insignificant for \kltomupm decay rate.
\par A second experimentally accessible quantity in \kltomupm decay is
the longitudinal polarization of the leptons $P_L$. The state $K_L$ is
a Superposition of the CP-eigenstates $K^0_1$ and $K^0_2$. 
\beq
\label{tone}
K_L = (1 + |\epsilon|^2)^{1/2}~~ (K^0_2 + \epsilon K^0_1)
\eeq
where we have followed the Wu-Yang phase convention and $\epsilon$ is
the CP - violating $K^0_1 - K^0_2$ mixing parameter given by: 
\beq
\label{ttwo} 
\epsilon \simeq  (2 \times 10^{-3}) ~~ exp(i \pi/4)
\eeq
It has been known
for a long time  \cite{sehgal2} that $P_L$ would be zero unless P and
CP are both violated in the decay. For the $K_L$ - decay, a finite
$P_L$ thus can arise {\it (i)} directly from CP - violating decay of
$K^0_2$ and {\it (ii)} from CP - conserving the decay of $K^0_1 \to
\mu^+ \mu^-$, because of the presence of $\epsilon K^0_1$ component of
$K_L$. The invariant amplitude for the decay $K^0_i (i = 1,2)$ into
$\mu^+(p_+) \mu^-(p_-)$ can be written as \cite{chang} :
\beq
\label{tthree}
{\cal M}_i = {\bar u}(p_-) ~[ a_i \gamma_5 + i b_i] ~v(p_+)
\eeq
where $a_2, b_1$ are CP - conserving and $a_1, b_2$ violate CP -
invariance. The amplitude for $K^0_L$ decay into $\mu^+ \mu^-$ is then
given by an expression similar to the above one with $a_i, b_i$
replaced by $a_L, b_L$ :
\beq
\label{tfour}
a_L = a_2 + \epsilon a_1 \quad, \quad b_L = b_2 + \epsilon b_1
\eeq
The longitudinal polarization $P_L$ can be expressed in terms of $a_2,
b_2$ as upto $O(\epsilon)$ 
\beq
\label{tfive}
P_L = \frac{m_K r^2 Im( a_2^* b_2 + a_2^* \epsilon b_1)}{4 \pi \Gamma} 
\eeq
where $r = (1 - 4 m_\mu^2/m_K^2)^{1/2}$, and $\Gamma$ is the total
decay width :
\beq
\label{tsix}
\Gamma = \frac{m_K r}{8 \pi} (|a_L|^2 + r^2 |b_L|^2)
\eeq
In the Standard Model (SM) the direct contribution proportional to
$Im(a_2^* b_2)$ in eqn.(\ref{tfive}), is small. The leading
contribution comes from the induce ${\bar s}d - {\rm Higgs}(H)$
vertex. This could potentially be large if the Higgs mass $m_H$ is
close to $m_K$ but in view of the current limits $m_H > 77.5 GeV$
\cite{rev} the direct contribution to $P_L$ would be smaller than
$10^{-4}$ \cite{botella}. The indirect contribution to $P_L$
represented by the term $Im(a_2^* \epsilon b_1)$ numerator of
eqn.(\ref{tfive}), has been investigated in detail by Herczeg
\cite{herczeg} assuming $a_2$ to be dominantly imaginary. A more
complete treatment without this assumption has been given by Ecker and
Pich \cite{ecker}. They obtain a value $|P_L|
\simeq 2 \times 10^{-3}$ but observe that in view of the uncertainties
of chiral perturbation theory employed for the estimate, experimental
value of $|P_L| > 5 \times 10^{-3}$ would be evidence for new physics
beyond SM.
\par The purpose of this note is to reexamine the direct contribution
to $P_L$ in the context of the minimal supersymmetric extension of the
standard model (MSSM) \cite{higgs}. Compared to SM, the parameter
space of MSSM is much bigger and the number of neutral Higgs jumps
from one in SM to three, two of which are CP - even and one CP - odd.
\par The immediate consequence of the minimum SUSY extension of the
standard model in respect of $\triangle S = 1$ neutral current
operators have been examined by Cho et.al. and Bertolini
et.al.\cite{cho,bertolini}. The basic structure of the effective
Hamiltonian for $\triangle S = 1 $ obviously remains unchanged since
the superpartner particles are all heavy and as with other heavy
particles do not make their appearance in the low-dimensional
operators responsible for low energy $\triangle 
S = 1$ processes. The effect of superpartner particles are felt
through their modifying the values of the various Wilson co-efficients
in the effective Hamiltonian :
\begin{eqnarray}
\label{tseven}
{\cal{H}}_{eff}
&=& 
\frac{\alpha G_F}{\sqrt{2} \pi} \lambda
\Bigg[
C^{eff}_9 ~({\bar s}~ \gamma^\mu~ P_L ~d)
({\bar l}~ \gamma^\mu ~l)
+ ~~C_{10}~({\bar s}~ \gamma^\mu~ P_L ~d)({\bar l} ~\gamma^\mu
~\gamma_5 ~l) \nonumber \\
&+& \frac{2 C_7 m_b}{p^2} ({\bar s}~ \not\!p ~\gamma^\mu ~P_R ~d)
( {\bar l}~\gamma^\mu ~\gamma_5~ l)
\Bigg]
\end{eqnarray}                     
with $p$ being the sum of the lepton momenta. The structure in
eqn. (\ref{tseven}) is obtained by taking account of box and
$Z^0$-penguin diagrams together with their superpartner
counterparts. Using the constraints of the MSSM parameter space forced
by experimentally observed $b \to s\gamma$ decay \cite{cleo}, the
changes in the Wilson co-efficients from their SM values have been
found to be mild. In any case the Hamiltonian eqn.(\ref{tseven})
results in a CP - invariant $K_L \to \mu^+ \mu^-$ amplitude and thus
does not contribute to $P_L$. However, if the parameter \tanbeta in
MSSM is large , of the order of 25 or more, the contribution of
neutral higgs bosons (NHBs) exchange amplitude (which is not included
in the effective hamiltonian eqn. (\ref{tseven})) can become
significant. The purpose of this note is to investigate this aspect. 
\par The dominant NHB exchange contributions to the effective
Hamiltonian for the process \kltomupm are shown in Figure 2. The
effective Hamiltonian from NHB has the structure \cite{huang,choudhury}:
\beq
\label{one}
{\cal H}_{eff}^{NHB} ~=~ \frac{\alpha G_F}{2 \sqrt{2} \pi} \lambda 
~\Bigg
[  C_{Q_1}~ {\bar s}(1 + \gamma_5) d ~{\bar l} l 
~+~  C_{Q_2}~ {\bar s}(1 + \gamma_5) d ~{\bar l} \gamma_5 l ~+~
h.c. \Bigg]
\eeq
$C_{Q_1}, C_{Q_2}$ are Wilson co-efficients at scale $\mu$, which for
our case will be $\sim 1 GeV$. The $C_{Q_1}$ term above contributes a
CP - violating piece to the invariant amplitude for \kltomupm :
\beq
\label{two}
{\cal M}^{NHB} ~=~ \frac{\alpha G_F}{2 \sqrt{2} \pi} ~
\frac{C_{Q_1}}{\sqrt{2}} ~2~ ( i ~Im \lambda) ~\langle ~0~ |~ {\bar s} \gamma_5
d ~ | ~K_0 ~\rangle ~{\bar u}(p_-) v(p_+)
\eeq
where $p_+, p_-$ are the momentum of $l^+$ and $l^-$ respectively. We
write the invariant amplitude following the convention of Herczeg
\cite{herczeg} :
\beq
\label{teight}
{\cal M} = a_2 {\bar u}(p_-) \gamma_5 v(p_+) + i b_2 {\bar u}(p_-)
v(p_+) 
\eeq
where the phases of $K_2$ has been chosen such that $a_2$ and $b_2$
are real except for unitary phases. In (\ref{teight}) we have taken the
$K_L$ amplitude to be the same as the CP - odd $K_2$, since we are
interested in the contribution to $P_L$ arising out of the direct
part. With this convention for the $K_L$ amplitude, we can relate the
matrix element of $({\bar s} \gamma_5 d)$ between vacuum and $K^0$ via
the kaon-decay constant ($f_K$) as follows :
\beq
\label{three}
\langle~ 0~ |~ {\bar s} \gamma^5 d ~|~ K_0~ \rangle ~=~ 
\frac{f_k m_k^2}{(m_s+ m_d)_c}
\eeq
where the suffix c indicates that the  masses are current quark
masses.
The NHB contributions to $b_2$ is : 
\beq
\label{five}
b_2^{NHB} ~=~ \frac{\alpha G_F}{2 \pi}~ C_{Q_1}(\mu) ~(Im \lambda)~
\frac{f_k m_K^2}{(m_s + m_d)_c}
\eeq
The amplitude $a_2$ is almost totally saturated by the $\gamma \gamma$
intermediate state where contribution has been estimated by Herczeg
\cite{herczeg} :
\beq
\label{six}
Im (a_2^{\gamma \gamma}) ~\approx ~2 \times 10^{-12}
\eeq
We shall use this value as the total $Im a_2$.We now then have all the
ingredients for estimating the NHB exchange graphs contribution to the
direct parts of the contribution to $P_L$  :
\beq
\label{seven}
P_L ~=~ \frac{2 r  ~Im ( b_2 a_2^*)}{ |a_2|^2 + (1 - \frac{4
m_\mu^2}{m_k^2})~|b_2|^2} 
\eeq
For numerical estimation, we use the value of $Im \lambda$ in terms of
Wolfenstein parameterization :
\beq
\label{eight}
Im \lambda ~=~ A^2 \lambda^5 \eta
\eeq
Using the input parameters as given in Appendix we get :
\beq
\label{nine}
P_L^{dir} ~=~ 0.4~ C_{Q_1}
\eeq
The numerical value expected thus is directly proportional to the
unknown Wilson co-efficient $C_{Q_1}(\mu)$ at scale $\sim$ 1 GeV. For
this corresponding co-efficient in $b \to s$ transition, this
co-efficient has been calculated in terms of MSSM parameters by Dai
et.al \cite{huang}. This was done in standard way, by calculating the
penguin terms perturbatively at scale $\sim M_W$ and then using
Renormalization Group equations (RGE) to evolve down to much lower
scales. The RGE evolution involves no operator mixing and so this is a
multiplicative correction in coming down from $M_W$ to $m_b$. For us,
the RGE is identical \& so the only difference will be in the
perturbative estimate of $C_{Q_1}(M_W)$. The mass of the quark enters
the calculation of this since the Higgs coupling to quarks is directly
proportional to the quark mass. Thus the $C_{Q_1}(\mu)$ for the $s \to
d$ transition will effectively be a factor $(m_s/m_b)$ down from its
value for the $b \to s$ transition. From a purely field theoretical
point of view, the masses above would be the masses in
the SM - langrangian, namely the 'current' quark masses, where values
for the light quark are determined through low-energy chiral symmetry
breaking analysis ( Cheng \& Li \cite{cheng} Section 5.5). We will use
the masses and Wolfenstein parameters of CKM as given in appendix.
\par The value of $C_{Q_1}$ depends crucially on MSSM parameters. As
is well known, MSSM has an undesirably long list of parameters. Most
phenomenological analysis in MSSM use unification model in which SUSY
is softly broken (at around Planck scale) leading to the 'SUGRA'
version of MSSM. Such models are completely specified by a common
gaugino mass term, a scalar mass term, trilinear coupling (all
specified at Planck scale) together with the higgs sector parameter and
\tanbeta in addition to SM parameters. Several authors \cite{lopez}
have analyzed this parameter space and the constraints imposed therein
by SM - parameters as well as by the now known $b \to s \gamma$ data as
given by CLEO \cite{cleo}. We in particular, work inside  
the parameter space as analyzed e.g. by Goto et.al \cite{goto} where
strict universality of the soft SUSY breaking mass holds separately
for squarks and scalars. With such relaxed universality, working
within allowed parameter space region consistent with all low energy
SM - parameters and $b \to s \gamma$, it is possible \cite{choudhury}
to have regions of parameter space where \tanbeta is large but the
NHBs are relatively light ( $ \approx 2 M_W$). Such allowed values of
MSSM parameters have a wide range wherein the $C_{Q_1}$ for $b \to s$
transition is of the order $O(1)$; the value of $C_{Q_1}$ for $s \to
d$ transition would be down by a factor $m_s/m_b \approx
0.025$. Fig. (3) shows typical values of the co-efficient $C_{Q_1}$,
for $s \to d$ transition relevant to \kltomupm decay. For value of the
CP-odd higgs mass ($m_A$) large (say greater than 200 GeV),
$C_{Q_1}$ indeed is too small . However, for
somewhat lower values of $m_A$, with $tan\beta > 25$ (which is within
the acceptable range of parameters), $C_{Q_1}$ can be sufficiently
large for the NHB - exchange contribution to $P_L$ (\ref{nine}) to
overwhelm the SM-estimate. Thus for a typically low value of $m_A =
150 GeV$, we get from eqn.(\ref{nine}) values of $P_L^{dir} = 0.7 \times
10^{-2}, 1.2 \times 10^{-2}, 1.8 \times 10^{-2}$ respectively for 
\tanbeta = 25, 30, 35.  

\par In summary, the predictions of MSSM for $P_L$ are as follows. If
the parameter \tanbeta is small then MSSM does not change the SM
value, which as stated before is dominated by the indirect
contribution and estimated at $|P_L| \simeq 2 \times 10^{-3}$ by
Ecker and Pich. For large \tanbeta and masses of Higgs bosons
exceeding 250 GeV, once again MSSM does not change SM
predictions. However if \tanbeta is large ( $\sim$ 25 or more) and the
Higgs masses are in the range of upto $2 M_W$ , the NHB-exchange
contributions to $P_L$ start becoming significant. A typical value
for \tanbeta = 30 and $m_A = 150 GeV$ gives $P_L = 1.2 \times 10^{-2}$.
When one is able to narrow down the acceptable parameter space of MSSM,
experimental measurement of $P_L$ would thus provide a very useful
confirmatory test.

\begin{center}
{\large \bf Acknowledgements} \\
\end{center}
One of the authors AG is thankful to CSIR for providing financial
support. 

\begin{center}
{\large \bf Appendix}  \\
Input parameters \\

Wolfenstein parameters \cite{kruger} ~:~~ $ A \cong 0.8 ~;~ \lambda = 0.22
~;~ \eta = 0.34 $ \\
Current quark masses \cite{cheng} ~:~~ $m_d = 7 ~MeV ~;~  m_s = 130~ MeV$ \\
$m_\mu = 105~ MeV ~,~ G_F = 1.16 \times 10^{-5}~ GeV^{-2} ~,~
\alpha_s(m_Z) = 0.119 ~,~ m_b \approx 5~ GeV$ 

\end{center}


\vfill

\pagebreak

\begin{center}
%
%

\begin{center}

\begin{picture}(400,500)(0,0)


\Line(0,480)(190,480)
\DashLine(30,480)(30,400){5}
\Line(30,400)(0,350)
\Line(30,400)(60,350)
\PhotonArc(120,480)(40,180,0){2}{6}

\Text(10,490)[c]{\large s}
\Text(120,490)[c]{\large ${\tilde q}$}
\Text(35,430)[l]{\large $h^0,H^0,A^0$}
\Text(180,490)[c]{\large d}
\Text(120,430)[c]{\large {$\tilde \chi$}}
\Text(0,340)[c]{\large $\mu^+$} 
\Text(60,340)[c]{\large $\mu^-$}

\Text(90,300)[c]{\Large (a)}


\Line(210,480)(400,480)
\PhotonArc(270,480)(40,180,0){2}{6}
\DashLine(360,480)(360,400){5}
\Line(360,400)(330,350)
\Line(360,400)(390,350)

\Text(220,490)[c]{\large s}
\Text(270,490)[c]{\large ${\tilde q}$}
\Text(355,430)[r]{\large $h^0,H^0,A^0$} 
\Text(270,430)[c]{\large ${\tilde \chi}$} 
\Text(390,490)[c]{\large d}
\Text(330,340)[c]{\large $\mu^+$}
\Text(390,340)[c]{\large $\mu^-$} 

\Text(300,300)[c]{\Large (b)}


\Line(0,200)(190,200)
\PhotonArc(95,200)(40,0,180){3}{6}
\DashLine(95,200)(95,120){5}
\Line(95,120)(65,70)
\Line(95,120)(125,70)

\Text(10,210)[c]{\large s}
\Text(70,190)[l]{\large ${\tilde q}$} 
\Text(95,260)[c]{\large ${\tilde \chi}$} 
\Text(180,210)[c]{\large d}
\Text(100,160)[l]{\large $h^0,H^0,A^0$}
\Text(65,60)[c]{\large $\mu^+$}
\Text(125,60)[c]{\large $\mu^-$}

\Text(90,20)[c]{\Large (c)}


\Line(210,200)(400,200)
\PhotonArc(305,200)(40,180,0){3}{6}
\DashLine(305,160)(305,100){5}
\Line(305,100)(275,50)
\Line(305,100)(335,50)

\Text(220,210)[c]{\large s}
\Text(305,210)[c]{\large ${\tilde q}$}
\Text(340,160)[c]{\large ${\tilde \chi}$} 
\Text(300,130)[r]{\large $h^0,H^0,A^0$} 
\Text(275,40)[c]{\large $\mu^+$}
\Text(335,40)[c]{\large $\mu^-$} 

\Text(300,20)[c]{\Large (d)}
\end{picture}

\end{center}


\end{center}
\noindent Figure~2:{\it The dominant contributions at scale $M_W$, of
NHB exchange contribution to the effective Hamiltonian for \kltomupm}
\vskip 1cm                                           

\pagebreak

\begin{center}
\epsfig{file=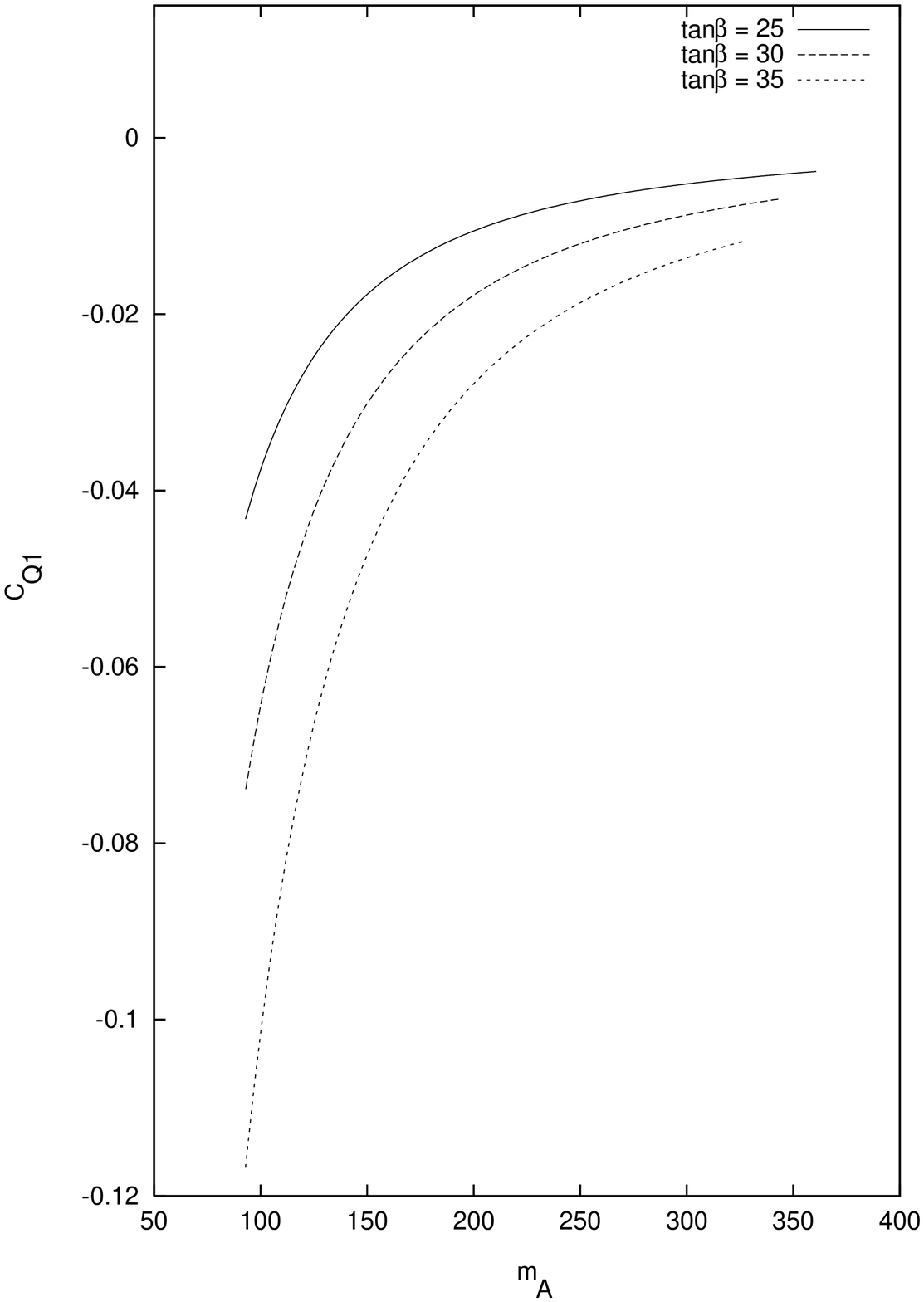,width=16cm,height=17cm} \\
\end{center}
\vskip 0.5cm
\noindent {Figure~3 ~:~{\it Typical values of $C_{Q_1}$ for $s \to d$
transition at GeV scale. Values of $C_{Q_1}$ 
have been plotted with pseudoscalar higgs mass $(m_A)$. The other MSSM
parameters are $m = M = 150~ GeV ~,~ A = -0.5$.}


\begin{thebibliography}{99}


\bibitem{buchalla1} For a review see G.Buchalla, A. Buras
and M.Lautenbacher {\em Rev. of Mod. Phys.}{\bf 68},1125 (1996).  
\bibitem{stone1} For a review see the article by A.Ali in B-decays
(revised $2^{nd}$ edition , World Scientific (Ed. S.Stone) 1994.
\bibitem{buras1} A.Buras and M.Munz \prd{52}{156}{1995}.
\bibitem{sehgal1} L.M.Sehgal {\em Nuovo Cimento}{\bf 45},785 (1963) ;
{\em Phys. Rev.}{\bf 183},1511(1969).
\bibitem{herczeg} P.Herczeg \prd{27}{1512}{1982} ; C.Q.Geng and J.Ng
\prd{41}{2351}{1990} ; G.Belanger and C.Q.Geng \prd{43}{140}{1991} ;
P.Ko \prd{45}{174}{1992}.
\bibitem{buchalla2} G.Buchalla, A.J.Buras \nuphsb{412}{106}{1994}.
\bibitem{sehgal2} L.M.Sehgal {\em Phys. Rev.}{\bf 181},2151 (1969)
\bibitem{chang} D.Chang and R.N.Mohapatra \prd{30}{2005}{1984} ;
J.Ellis,M.K. Gaillard and D.V.Nanopoulos \nuphsb{106}{292}{1976};
R.S.Willey and H.L.Yu \prd{26}{3086}{1982}; B.Grzadkowski and
P.Krawczyk {\em Z.Phys.}{\bf C 18},43 (1983).
\bibitem{botella} F.J.Botella and C.S.Lim \prd{56}{1651}{1986}.
\bibitem{rev}Particle Data Group,{\em The European Physical Journal}
{\bf C3} (1998)1.         
\bibitem{ecker} G.Ecker and A.Pich \nuphsb{366}{189}{1991}.
\bibitem{higgs} For a review of the structure of MSSM see ``The Higgs
Hunters Guide'' : J.Gunion, H.Haber, G.Kane \&
S.Dawson. Addison-Wesley Publishing Co.(1996).
\bibitem{cho} P.Cho, D. Wyler and M. Misiak \prd{54}{3329}{1996}.
\bibitem{bertolini} S. Bertolini et.al. \nuphsb{353}{591}{1991}.
\bibitem{cleo} CLEO Collaboration,M.S.Alam et.al. Phys.Rev.Lett.~
{\bf 74} , 2885 (1995); CLEO Collaboration R.Ammar et.al, ibid. {\bf
71}, 674 (1993).    
\bibitem{kruger} F. Kruger and L.M.Sehgal \prd{55}{2799}{1997}.
\bibitem{cheng} ``Gauge theories of Elementary particle Physics''
T.P.Cheng and L.F.Li; Clareden Press, Oxford (1988).
\bibitem{huang} C-S.Huang,W.Liao and Q-S.Yan \prd{59}{011701}{1999};
Y-B.Dai et.al. \plb{390}{257}{1997}; C-S.Huang and Qi-Shu Yan
\plb{442}{209}{1998}. 
\bibitem{choudhury} S.Rai Choudhury and Naveen Gaur
\plb{451}{86}{1999}; S.Rai Choudhury,Naveen Gaur
and Abhinav Gupta \prd{60}{115004}{1999}.
\bibitem{goto} T.Goto, Y. Okada and Y. Shimuzu Phys. Rev {\bf D 58},
094006 (1998).           
\bibitem{lopez}J.L.Lopez, D.V.Nanopoulos, Xu Wang and A.Zichichi
Phys.Rev. {\bf D 51}, 147 (1995);  J.L.Lopez, D.V.Nanopoulos and
A.Zichichi Phys. Rev. {\bf D 49}, 343 (1994).       


\end{thebibliography}
\end{document}